\documentclass[RNAAS]{aastex62}
\usepackage{CJKutf8}

\graphicspath{{./}{figures/}}

\begin{document}

\title{A Revised Velocity for the Globular Cluster GC-98 in the Ultra Diffuse Galaxy NGC\,1052-DF2}

\correspondingauthor{Pieter van Dokkum}
\email{pieter.vandokkum@yale.edu}

\author{Pieter van Dokkum}
\affiliation{Yale University}

\author{Yotam Cohen}
\affiliation{Yale University}

\author{Shany Danieli}
\affiliation{Yale University}

\author{Aaron Romanowsky}
\affiliation{San Jos\'e State University}
\affiliation{University of California Observatories}

\author{Roberto Abraham}
\affiliation{University of Toronto}

\author{Jean Brodie}
\affiliation{University of California Observatories}

\author{Charlie Conroy}
\affiliation{Harvard-Smithsonian Center for Astrophysics}

\author{J.~M.\ Diederik Kruijssen}
\affiliation{Universit\"at Heidelberg}

\author{Deborah Lokhorst}
\affiliation{University of Toronto}

\author{Allison Merritt}
\affiliation{MPIA, Heidelberg}

\author{Lamiya Mowla}
\affiliation{Yale University}

\author{Jielai Zhang}
\affiliation{University of Toronto}


\gdef\blob{NGC\,1052-DF2}
\section{} 
We recently published velocity measurements of luminous globular clusters
in the galaxy NGC\,1052-DF2, concluding that it lies far off the canonical stellar mass\,\,--\,\,halo mass (SMHM) relation \citep{vd18a} [vD18]. Here we present a revised velocity for one of the globular clusters, GC-98. 

\section{Lost and Found: the LRIS spectrum of GC-98}

GC-98 has been the subject of some debate. In vD18 we listed its velocity as $cz=1764^{+11}_{-14}$\,km/s, 39\,km/s removed from the central value of the sample. We inferred that the intrinsic velocity distribution of \blob\ is consistent with a Gaussian only for the narrow range $8.8<\sigma_{\rm intr}<10.5$\,km/s (at 90\,\% confidence), with the lower limit driven by GC-98. \citet{m18} [M18] derive a 90\,\% upper limit of $\sigma_{\rm{}intr}<17.3$\,km/s if GC-98 is included and $\sigma_{\rm{}intr}<14.3$\,km/s if it is not.
Although all these estimates imply a strongly dark matter-deficient galaxy (see Discussion), the quantitative constraints on the halo mass are sensitive to the treatment of this single object.

As discussed in vD18 most of the clusters were observed using two different spectrographs (DEIMOS and LRIS on Keck). We thought we had observed GC-98 only once, with DEIMOS. However, we recently realized that we also observed the object in one of the LRIS masks. The combined 28,800\,s LRIS+DEIMOS spectrum is shown in Fig.\,\,1. Using the methodology described in vD18 we derive a radial velocity of $cz=1784^{+10}_{-10}$\,km/s. The large change can be attributed to systematic residuals in the fit to the DEIMOS spectrum (see vD18, Fig.\,\,2).

\section{Revised Velocity Dispersion}

The velocities of the 10 clusters are shown in Fig.\,\,1, ordered by their absolute distance from the mean. The red curve is the expected distribution if the galaxy has very little dark matter; this is approximated by a Gaussian with $\sigma_{\rm stars}=7.0^{+1.6}_{-1.3}$\,km/s perturbed by the errors. We determined $\sigma_{\rm stars}$ using the \citet{wolf} relation with $M_{\rm stars}=2.0^{+1.0}_{-0.7}\times{}10^8$\,M$_{\odot}$ and $R_{\rm e}=2.2$\,kpc. The black curve is the expected dispersion of a halo with mass $6\times{}10^{10}$\,M$_{\odot}$, assuming that $\sigma_{\rm{}ap}(0.1R_{\rm{}v})=(0.6\pm{}0.1)V_{\rm{}v}$ \citep[see][]{lokas,smhm}.

In the right panel we show the constraints on the intrinsic dispersion, using two methods: the likelihood and Approximate Bayesian Computation \citep[ABC;][]{abc}.  The main advantage of ABC is that it does not assume a particular form of the likelihood function. In vD18 we used the biweight scale as the ABC summary statistic; here we simply use the rms. We derive a revised dispersion of $\sigma_{\rm{}intr}=5.6^{+5.2}_{-3.8}$\,km/s ($\sigma_{\rm{}intr}<12.4$\,km/s at 90\,\% confidence). The likelihood gives $\sigma_{\rm{}intr}=7.8^{+5.2}_{-2.2}$\,km/s ($\sigma_{\rm{}intr}<14.6$\,km/s).


\begin{figure}[htbp]
\begin{center}
\includegraphics[scale=0.95,angle=0]{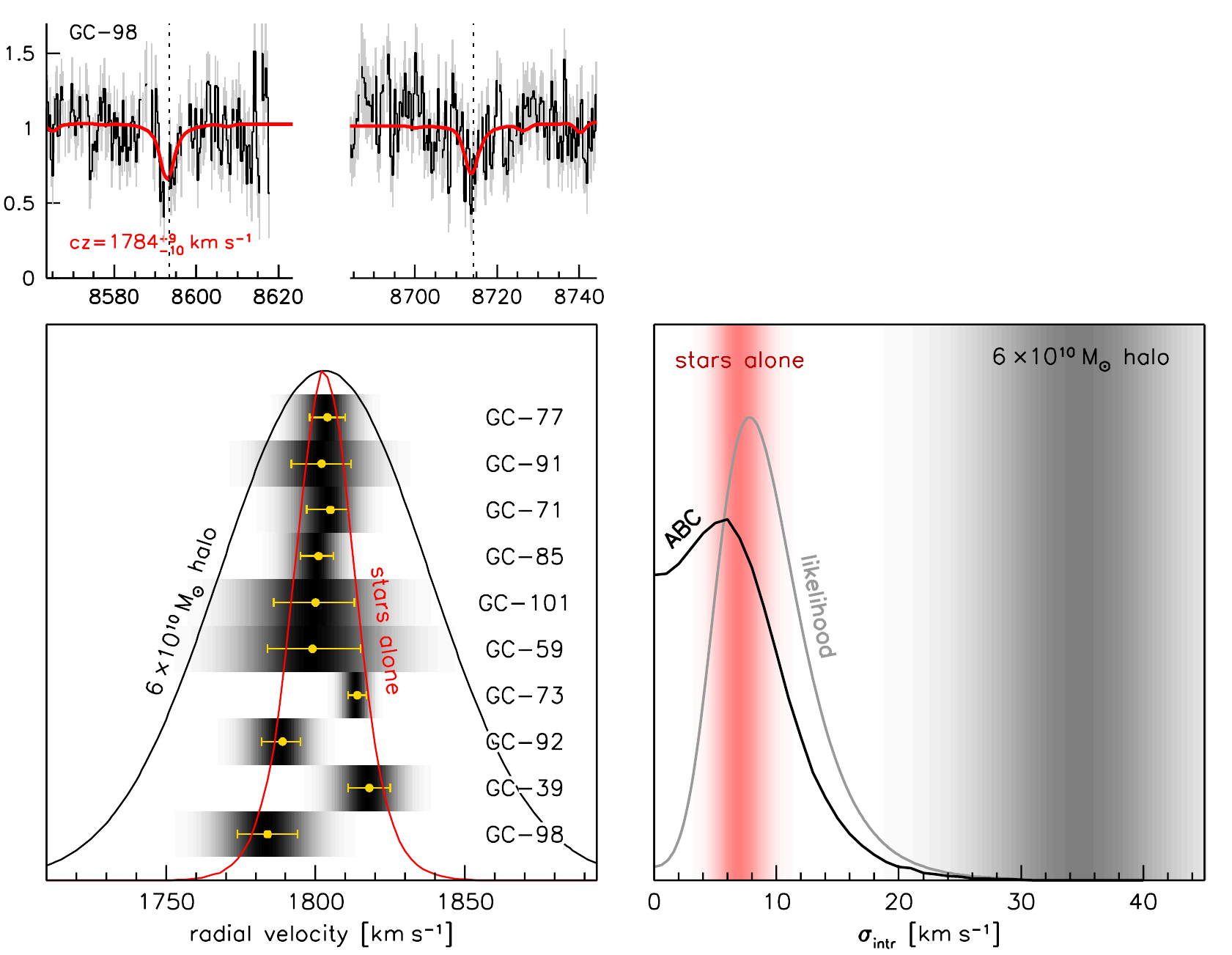}
\caption{Top: combined DEIMOS+LRIS spectrum of GC-98. Left: velocities of the 10 globular cluster-like objects in \blob.
Right: constraints on the intrinsic velocity dispersion. We find $\sigma_{\rm{}intr}=5.6^{+5.2}_{-3.8}$\,km/s. The stars alone contribute $\sigma_{\rm{}stars}=7.0^{+1.6}_{-1.3}$\,km/s; the expectation from the SMHM relation is $\sigma_{\rm{}intr}=35\pm{}6$\,km/s.}
\end{center}
\end{figure}

\section{Discussion}

The revised velocity dispersion is nearly identical to that expected from the stars alone, and does not significantly alter the analysis of vD18. The implied ratio $M_{\rm{}halo}/M_{\rm{}stars}$ is of order unity and consistent with zero. The expectation from the SMHM relation is $M_{\rm{}halo}/M_{\rm{}stars}\sim{}300$.

M18 suggest that \blob\ might not be dark matter deficient, based on the  upper limit that they derive on its $M/L$ ratio. Specifically, M18 infer that $M/L$ could be as high as 8.1 within $R=7.6$\,kpc, from their 90\,\% upper limit on the dispersion ($\sigma_{\rm{}intr}<17.3$\,km/s). However, $M/L$ is a strong function of radius, and we caution against using this quantity by itself as a proxy for halo mass.
An $M/L$ ratio $<8.1$ implies a dark matter mass within $R=7.6$\,kpc of $<6\times{}10^8$\,M$_{\odot}$, and a halo mass $<1\times{}10^9$\,M$_{\odot}$ \citep[see also][]{laporte}. As the expected halo mass from the SMHM relation is $\sim{}6\times{}10^{10}$\,M$_{\odot}$, {\em the M18 analysis implies that \blob\ lies at least a factor of 60 below the SMHM relation}. Because of their large spatial extent for their luminosity, the expected $M/L$ ratio for UDGs on the SMHM relation is of order 30--100, as observed in Virgo and Coma \citep[see Fig.\,\,4 in][]{toloba}.

In summary, the existence of \blob\ adds to the evidence that the SMHM relation has considerable scatter at the low mass end \citep[e.g.,][]{m14,oman}. It is impossible to say whether the galaxy has no dark matter at all but it is clearly extremely dark matter deficient. Next we aim to measure the {\em stellar} velocity dispersion of \blob, to test whether the velocity distribution of the globular clusters is approximately isotropic.  Finally, we emphasize that the kinematics and size are not the only remarkable aspects of \blob, as its globular cluster population is different from all other known galaxies \citep[see][]{vd18b}. It seems likely that the unusual properties of \blob\ have a common origin, but what this might be is still elusive.




\acknowledgments
We thank the team from
\begin{CJK}{UTF8}{min}日本放送協会\end{CJK}(NHK) for their invaluable contribution to the ``discovery" of the LRIS spectrum.

\end{document}